\documentclass{elsarticle}
\usepackage{amsmath}
\usepackage{amsfonts}
\usepackage{amssymb}
\usepackage{graphics}
\usepackage{graphicx}
\usepackage{dsfont}
\usepackage{setspace}
\usepackage{url}
\usepackage{fancybox,framed}
\usepackage[latin1]{inputenc}

\usepackage{setspace}
\usepackage{color}
\usepackage{xcolor}
\usepackage{hyperref}

\def\proof{{\noindent \textsc{Proof}. \;}}
\def\qed{$\Box$}

\newtheorem{theorem}{\bf Theorem}
\newtheorem{proposition}{\bf Proposition}
\newtheorem{remark}{\bf Remark}
\newtheorem{definition}{\bf Definition}
\newtheorem{example}{\bf Example}

\makeatletter
\def\ps@pprintTitle{%
  \let\@oddhead\@empty
  \let\@evenhead\@empty
  \def\@oddfoot{\reset@font\hfil\thepage\hfil}
  \let\@evenfoot\@oddfoot
}
\makeatother

\begin{document}
\onehalfspacing
\parskip=1mm
\begin{frontmatter}

\title{Two characterizations of the dense rank}
\author{Jos\'e Luis Garc\'{\i}a-Lapresta, Miguel Mart\'{\i}nez-Panero$^*$}\ead{\{lapresta,miguel.mpanero\}@uva.es}
\address{IMUVA, PRESAD Research Group, Departamento de Econom\'{\i}a Aplicada, \\
Universidad de Valladolid, Valladolid, Spain\\
$^*$Corresponding author}

\begin{abstract}
In this paper, we have considered the dense rank for assigning positions to alternatives in weak orders. If we arrange the alternatives in tiers (i.e., indifference classes), the dense rank assigns position 1 to all the alternatives in the top tier, 2 to all the alternatives in the second tier, and so on. We have proposed a formal framework to analyze the dense rank when compared to other well-known position operators such as the standard, modified and fractional ranks.
As the main results, we have provided two different axiomatic characterizations which determine the dense rank by considering position invariance conditions along horizontal extensions (duplication), as well as through vertical reductions and movements (truncation, and upwards or downwards independency).

\medbreak
\end{abstract}
\end{frontmatter}

\section{Introduction} \label{sect:introduction}
When it is possible to rank order objects (individuals, alternatives, etc.) taking into account some quality or criterion, it is natural to assign positive integer numbers to them in an ascending manner, starting from 1 for the best, 2 for the following one, and so on.
Such numbers can be interpreted as rankings\footnote{It is extended the double use of \emph{ranking} for both the preference relation on a set of objects and for the (ranking) numbers assigned to such objects. In the last case, along this paper we will use the term \emph{positions}.} or positions in an ordinal sense (first, second, etc.).
According to Kendall~\cite[p. 1]{Kendall-1948}, at first sight they might seem not allow computations:
``We cannot substract `fifth' from `eight'; but a meaning can be given  [taking into account that] to say that the rank according to A is 5 is equivalent to saying that [...] four members are given priority over our particular member, or are \textit{preferred} to it; [...] this is not an ordinal number but a cardinal number, i.e., arises by counting".
Italics (author's own) suggest the formal use of preference relations in the process of assigning positions, which is just the main technical tool in our approach along this paper.

However, when comparing alternatives ties can arise for several reasons, among them:
imprecision or lack of knowledge in agents' assessing process or, in some contexts, exact coincidence in the considered quality to be evaluated.
In such cases, positions to be assigned are not straightforward and might depend on the particular scenario.

For example, in the 2020 Olympic Games (held at Tokyo in 2021 due to the COVID pandemic), the Men's High Jump competition produced a strict tie, both in attempts and top exceeded height, between two athletes.
The rules  established either a tie-break (jump off) or to assign a shared award in equal terms by consensus agreement of the involved jumpers, and this last option was chosen.
In this way, this discipline had two gold medals, no silver and one bronze\footnote{See the official schedule at \url{https://olympics.com/en/olympic-games/tokyo-2020/results/athletics/men-s-high-jump}, and more details at \url{https://en.wikipedia.org/wiki/Athletics_at_the_2020_Summer_Olympics_Men's_high_jump}.
On the jump off and how it became optional after 2009, allowing this shared gold,  see \url{https://en.wikipedia.org/wiki/High_jump}.}.
This $1-1-3$ way of assigning positions, known as the \textit{standard (competition) rank}, is just one of the possibilities, but not precisely that our paper is focused on.

Should the jumper currently awarded with the bronze medal have been promoted to silver after the agreement between the two \textit{ex aequo} gold medals?
Would have been the positions to be assigned $\,1-1-2$, without a jump between consecutive positions, instead of the actual ones?
Not perhaps in this context of sports international prestige or whenever awards have a consequent monetary reward.
However, such approach (ours in this paper) makes sense in some other situations.

Consider a process where people can ask for something via the Internet and make accidental mistakes by multiple clicking, or even try strategically to  maximize their opportunities (by bots, for example).
If the enabled service detects that, from the same origin and in a few time, multiple requests are received, it would seem reasonable to identify them as just one and to assign one single request\footnote{This example is related to \textit{false-name-proof mechanisms}, i.e., those where individuals do not gain an advantage from participating more than once. An overview of false-name manipulation in several contexts (voting, auctions, etc.), and mechanisms to prevent it, can be found in Conitzer and Yokoo~\cite{Conitzer-Yokoo-2010}).},
say $\,n-n-n- \cdots -n\,$ ($m$ times), so that the following different petitioner will receive the next list number, $\,n+1\,$ (instead of $n+m$).

This way of assigning positions takes into account in the first instance not alternatives but ranking levels (gathering indifferent alternatives).
Commonly known as \emph{dense rank}\footnote{Because, as pointed out by Kyte~\cite[p. 563]{Kyte-2005} ``a dense rank returns a ranking number without any gaps".
More seldom, it is also called \emph{sequential rank}.}, it is widespread used as a convenient position method in some contexts. For example, the following question taken from Kyte~\cite[pp. 562-568]{Kyte-2005}, shows how the dense rank naturally appears in a company management scenario where just the rank would not be suitable enough\footnote{See \url{https://asktom.oracle.com/pls/apex/asktom.search?tag=dense-rank-vs-rank}.}.
``Consider this seemingly sensible request: Give me the set of sales people who make the top 3 salaries, that is, find the set of distinct salary amounts, sort them, take the largest three, and give me everyone who makes one of those values. [...]
We can simply select all [the individuals] with a dense rank of three or less. [...]
In this case, using [standard] rank over dense rank would not have answered our specific query".
This happens because the dense rank primarily focuses on obtained salaries, scores, etc., and afterwards on individuals who reach them.

From a theoretical point of view, the dense rank has received less attention than the standard rank (the current case of the Olympic medals) or the \textit{fractional rank} (also called \emph{mid-rank}), where the average of the positions as if there were no ties would be computed, and assigned to the alternatives in the tie: $1.5 - 1.5 - 3$, with a kind of gold-silver medal at 50\% alloy if applied for the \textit{ex aequo} Olympic winners).
Even more, another possibility would exist by taking $2-2-3$, without gold medal, that corresponds to the \textit{modified (competition) rank}.

In Table~\ref{tab:ranks}, the positions of the above Olympic example are shown for all considered ranks.

\begin{table}[h]
\centering
\begin{tabular}{c|cccc}
& Standard rank & Modified rank & Fractional rank & Dense rank\\
\hline
&&&&\\[-1.5ex]
$x\quad y$  & 1 & 2 & 1.5 & 1 \\[1ex]
$z$ & 3 &  3 & 3 &2 \\[1ex]
\hline
\end{tabular}
\caption{Ranks.}
\label{tab:ranks}
\end{table}

We have intentionally avoided before, and also will do in the sequel, the so called \emph{ordinal rank}, which assigns distinct positions to different objects, even when they are in a tie (for instance, with a random tie-breaking process).
In the previous example, although $x$ and $y$ had an equal performance over that of $z$, the ordinal rank could assign positions $\,1-2-3\,$ or $\,2-1-3\,$ to the alternatives $x,y,z$.
The reason for discarding these possibilities is that \emph{equality}, a compelling property to appear formally in Definition~\ref{def:position operator properties}, is vulnerated.

An interesting overview relating the dense rank to these and other ranking methods, with a suitable mathematical treatment, can be found in Vojnovi\'{c}~\cite[pp. 505-506]{Vojnovic-2016}.

In the literature, the dense rank is not considered at all by Kendall~\cite[pp. 34-48]{Kendall-1945,Kendall-1948} when focusing on ties.
It is, in passing and tacitly, by Fishburn~\cite[pp. 164-165]{Fishburn-1973}, related to his \emph{modified equal-spacing procedure};
and it is equivalent to the so-called \emph{ranking level function} introduced by G\"{a}rdenfors~\cite{Gardenfors-1973} in his analysis of positional voting functions.
More recently, in a similar way, Ding \emph{et al.}~\cite{Ding-Han-Dezert-Yang-2018} have proposed a novel method which gathers ambiguous alternatives at different stages in a hierarchical process of aggregation fusion.

Concerning software implementation, although both standard rank and  fractional rank exist as \texttt{EXCEL} functions, dense rank does not.
But it appears in \texttt{SPSS} as the rank case \texttt{SEQUENTIAL RANKS TO UNIQUE VALUES}\footnote{See \url{https://www.ibm.com/docs/en/spss-statistics/29.0.0?topic=cases-rank-ties}.}, and in \texttt{SQL} it is an analytical (or window) function called, precisely, \texttt{DENSE\_RANK}\footnote{See Kyte~\cite[pp. 562-568]{Kyte-2005}.}.
As a programming language for storing and processing information, \texttt{SQL} is supported by Amazon Web Services (AWS), where it is also possible to find applications of \texttt{DENSE\_RANK} to the logistics industry, such as the arrangement of products in inventory according to their quantities in order to prioritize restocks\footnote{See \url{https://docs.aws.amazon.com/redshift/latest/dg/r_WF_DENSE_RANK.html}.}.

It is important to emphasize that the four considered position operators (standard, modified, fractional and dense ranks) are equivalent from and ordinal point of view.
However, from a cardinal approach, they assign positions to the alternatives in a different fashion, still representing the same weak order.
This fact crucially affects the results when the positions are the inputs of an aggregation procedure, because they could generate different outcomes\footnote{Following with the Olympic example, the most extended way to establish an international overall ranking is a lexicographic order: first, the total amount of gold medals for each country is taken into account; then, the number of silver medals for breaking possible ties; and, finally, the bronzes, if necessary.
This approach only considers ordinal information from the distinct sport disciplines.
But there are other possibilities which use cardinal information from the same ordinal basis, by translating  positions into scores.
When this option is followed, the international ranking would depend on the use of standard, modified, fractional or dense ranks and their respective scoring conversions.}.
For instance, this is the case of consensus measures in the setting of weak orders.
In this way, Garc\'{\i}a-Lapresta and P\'erez-Rom\'an~\cite{Lapresta-Roman-book} consider the positions through the fractional rank.
In turn, Alcantud et al.~\cite{Alcantud-Andres-Gonzalez-2013} and Gonz\' alez-Arteaga et al.~\cite{Gonzalez-Alcantud-Andres-2016} propose some consensus measures based on the standard rank.
And, as expected, different ways of assigning positions to the alternatives may lead to divergent results when consensus is measured (see Alcantud et al.~\cite[Ex. 3.7]{Alcantud-Andres-Gonzalez-2013}).

In a different setting, the rank correlation coefficients introduced by Spearman and Kendall use the fractional rank, although it would be possible to change it by using other positions operators, as the standard, modified of dense ranks\footnote{Kendall \cite{Kendall-1945} himself explicitly considered the use of the standard rank (suggested by Student), but pointed out that ``it gives different results if one ranks from the other end of the scale and [...] destroys the useful property that the mean rank of the whole series shall be $\,\frac 12 (n+1)$".
This symmetry is the main reason for the extended use of the fractional rank (mid-rank) in correlation analysis.}.
Obviously, as happens in consensus measures, the outcomes would be different depending on the used rank.

In the frameworks of positional voting systems (see G\"{a}rdenfors~\cite{Gardenfors-1973} and Garc\'{\i}a-Lapresta and Mart\'{\i}nez-Panero~\cite{Lapresta-Panero-2017}) and scoring rules (see Chevotarev and Shamis~\cite{Chevotarev-Shamis-1998}), when these voting systems are extended from linear orders to weak orders, the way that positions are defined is crucial to generate a collective ranking of the alternatives.
For example,  Madani et al.~\cite{Madani-Read-Shalikarian-2014} show how ranking orders (outputs) are affected by the position operator chosen to aggregate the inputs.
Although the fractional rank is commonly used in the most well-known scoring rule, the Borda count (see Black~\cite{Black-1976} and Cook and Seiford~\cite{Cook-Seiford-1982}, among others),
G\"{a}rdenfors~\cite{Gardenfors-1973} considers other variants and he also defines the \emph{restricted Borda function}, based on the modified rank, and (as mentioned above) the \emph{ranking level function}, related to the dense rank.

It is not the objective of this paper to decide which position operator is the most suitable one in the mentioned and other contexts.
Here we pretend to discriminate what are the intrinsic positional properties of the dense rank, in particular those not fulfilled by the other usual considered position operators.
To this aim, this paper proposes a first characterization of the dense rank as the only position operator verifying two independent conditions:
sequentiality (the positions of alternatives in linear orders should follow the natural sequential pattern: $\,1, \, 2,\, 3,\,\dots$); and duplication, a kind of ``clone\footnote{Cloning plays a relevant role not only in Genetics, but also in other scientific fields, such as Quantum Physics and Voting Theory, being this last approach the closest to ours.
In Section~\ref{sect:characterizations} we extensively explain this affinity.}
independency", meaning that the appearance of new replicated alternatives ought not modify the positions of all already existing ones, at any level.

A second axiomatization is also provided substituting duplication by a condition called UD-independency (upwards or downwards independency).
While duplication entails the preservation of the original positions by cloning (i.e., addition of indifferent alternatives to the already existing ones, in a horizontal way), UD-independency states that vertical displacements of alternatives from the original situation will not change the positions of the remaining non moving alternatives\footnote{This requirement makes sense in some contexts (related to those in which duplication also does) where one's improvement or setback do not affect to others.
Again, in Section~\ref{sect:characterizations} we detail this.}.
Being this condition weaker than duplication (a non immediately evident fact due to the distinct scopes of both properties), in order to achieve a second characterization of the dense rank we will need to add truncation (a compelling condition meaning irrelevance of alternatives below when assigning positions) and, again, as in the first characterization, sequentiality.

The rest of the paper is organized as follows.
Section~\ref{sect:dense rank} introduces the notation and the new framework of position operators to represent the dense rank in a formal way within the setting of weak orders.
Section~\ref{sect:basic properties} is devoted to explain basic properties required to position operators.
Section~\ref{sect:characterizations} includes other different properties, their relationships with the basic ones and the characterization theorems of the dense rank.
Section~\ref{sect:concluding remarks} concludes with some remarks, paying special attention to some differences among dense, standard, modified and fractional ranks, and showing how the dense rank is essentially different to the others.

\section{The dense rank in the setting of weak orders} \label{sect:dense rank}
In this section we give an in-depth outlook of ties and the problem they present when assigning positions, and focus on the dense rank as a way of deal with this situation.
To this aim, first of all we provide some notation used throughout the paper.

\subsection{Notation}
Consider a finite set of alternatives $\,X=\{x_1,\dots,x_n\}$, with $\,n\ge 2$.
A \emph{weak order} (or \emph{complete preorder}) on $X$ is a complete and transitive binary relation on $X$.
A \emph{linear order} on $X$ is an antisymmetric weak order on $X$.
With $\,\mathcal{W}(X)\,$ and $\,\mathcal{L}(X)\,$ we denote the sets of weak and linear orders on $X$, respectively.
Given $\,R\in \mathcal{W}(X)$, with $\,P\,$ and $\,I\,$ we denote the asymmetric and the symmetric parts of $R$, respectively:
$\,x_i \,P\, x_j\,$ if not $\,x_j\,R\,x_i$; and $\,x_i \,I\, x_j\,$ if ($x_i\,R\,x_j\,$ and $\,x_j\,R\,x_i$).

Given $\,R\in \mathcal{W}(X)\,$ and a permutation $\sigma$ on $\,\{1,\dots,n\}$, we denote by $\,R^{\sigma}\,$ the weak order obtained from $R$ by relabelling the alternatives according to $\sigma$, i.e., $\,x_i\,R\,x_j \;\Leftrightarrow\; x_{\sigma(i)}\,R^{\sigma}\,x_{\sigma(j)}$, for all $\,i,j \in \{1,\dots,n\}$. As usual, we denote by $(i,j)$ the transposition interchanging these two subindexes and keeping unaltered all others.

Given $\,R\in \mathcal{W}(X)\,$ and $\,Y \subseteq X$, the restriction of $R$ to $Y$, $\,R\vert_Y$, is defined as $\,x_i \,R\vert_Y \,x_j$ if $\,x_i \,R\, x_j$, for all $\,x_i,x_j \in Y$.
Note that $\,R\vert_Y \in \mathcal{W}(Y)$.

In turn, $\,\# Y$ is the cardinality of $Y$.

\subsection{Positions in weak orders}
We now introduce the notion of position operator, a suitable mathematical object which allows us to add or withdraw alternatives along the process of assigning positions, in a parallel way to that of voting theory when a variable electorate is considered (see~Smith \cite{Smith-1973}).

\begin{definition} \label{def:position operator}
Given a universe of alternatives $\,U$ and $\,X \subseteq U\,$ finite, a \emph{position operator} $O$ assigns to each $\,R\in \mathcal{W}(X)\,$ a function $\,O_R: X \longrightarrow \mathbb{R}$.
We say that $\,O_R (x_i)\,$ is the \emph{position} of the alternative $\,x_i \in X\,$ in the weak order $R$.
\end{definition}

In Definitions~\ref{def:position operator properties}, \ref{def:duplication} and \ref{def:UD-i} we include some properties that position operators may fulfil.

Once $X$ has been fixed, given $\,R\in \mathcal{W}(X)\,$ and $\,x_i\in X$, with $\,p_i\,$ we denote the number of alternatives dominated by $x_i$:
$$
p_i= \# \left\{x_j\in X \mid x_i \,P\, x_j\right\}.
$$

Note that $\,p_i\in \{0,1,\dots,n-1\}$.

Assigning positions to the alternatives in a linear order is a trivial task, as shown in the following definition.

\begin{definition} \label{def:sequential function}
Given $\,R\in \mathcal{L}(X)$, the \emph{sequential function} on $R$ is the mapping $\,S_R:X \longrightarrow \{1,\dots,n\}\,$ that assigns 1 to the alternative ranked first, 2 to the alternative ranked second, and so on:
$$
S_R(x_i) = \#\left\{x_j\in X \mid x_j \,R\, x_i\right\} = n - \#\left\{x_j\in X \mid x_i \,P\, x_j\right\} = n-p_i.
$$
\end{definition}

Differently to the case of linear orders, it is not obvious how to assign positions to the alternatives in weak orders, where ties may appear (we have already dealt with this fact and pointed out some possibilities in Section~\ref{sect:introduction}).

According to Kendall~\cite{Kendall-1945}, there exist different ways of ``allocating ranking numbers to tied individuals" [assigning positions to indifferent alternatives, in our context]:

\begin{itemize}
\item
``The method of allocating ranking numbers to tied individuals in general use is to average the ranks which they cover.
For instance, if the observer ties the third and fourth members, each is allotted $\,3\frac{1}{2}$.
This is known as the mid-rank method".

\item
``An alternative to mid-ranks [is] that the ties should all be ranked as if they were the highest member of the tie", i.e., 3 in the previous example.
This is the standard rank, the most common method where just natural numbers are used to assign positions.
This is the reason why it is often called simply ``the rank", avoiding the adjective ``standard".
\end{itemize}

And of course, although not considered by Kendall~\cite{Kendall-1945, Kendall-1948}, it is also possible to assign as position the lowest member in the tie (4 in Kendall's example).
This is called the modified rank.

Note that these three cases can be understood as follows.
By means of a tie-breaking process we can reduce the problem of assigning positions to the linear case, and then the average, highest or lowest values are given to the tied alternatives (the same for all of them\footnote{This desirable property will be called \emph{equality} from Definition~\ref{def:position operator properties} onwards.}), respectively.
Nonetheless, other possibilities exist, and we consider a different way, not appearing in Kendall~\cite{Kendall-1945, Kendall-1948} either, that is not defined through a tie-breaking procedure: the dense rank.

\subsection{Dense rank}  \label{subsect:dense rank}
 G\"{a}rdenfors~\cite{Gardenfors-1973} points out that ``two alternatives $x$ and $y$ are at the same ranking level in the preference order $R\,$ iff $\,x\,I\,y$.
 This is an equivalence relation and the equivalence classes are called ranking levels" (we use the visual term \textit{tiers}).
In other words, all \textit{ex aequo} classified alternatives (at any tier) ought to be treated in the same way, as just only one alternative (equality).

We now formally establish a linear order in the quotient set $X/I$, whose elements are the induced  tiers, and where positions are univocally given by the sequence of natural numbers.
And then we will extend this approach from $\,X/I\,$ to $X$, i.e., from tiers to alternatives in each tier, sharing the same positions.
Hence, the dense rank is a compelling ranking method which consists in assigning position 1 to the alternatives in the top tier, position 2 to the alternatives in the second tier, and so on.

In this way, consider again the above example taken from Kendall ~\cite{Kendall-1945}, where, among several individuals, there is a tie between the third and fourth members.
While the above considered positions range from 3 to 4 (depending on the ranking method used), their dense rank might even change to 2 if those other alternatives above them were also in a tie.

Next we introduce the notation in order to formally deal with the dense rank.

\begin{definition} \label{def:T_p and T}
Given $\,R\in \mathcal{W}(X)$, for each $\,p\in \{0,1,\dots,n-1\}$, with $\,T_p\,$ we denote the tier gathering all the alternatives that have $p$ alternatives below,
\begin{equation} \label{eq:Tp}
T_p=\{x_i\in X \mid p_i=p\},
\end{equation}
and
\begin{equation} \label{eq:T}
T=\big\{p \in \{0,1,\dots,n-1\} \mid T_p \ne \emptyset \big\}.
\end{equation}
\end{definition}

Hereinafter, when we say tiers, we mean non-empty tiers i.e., $\,T_p\,$ with $\,p\in T$.

\begin{remark} \rm
If $\,R\in \big(\mathcal{W}(X) \setminus \mathcal{L}(X)\big)$, some $T_p$ will be empty.
However, always $\,T_0 \ne \emptyset$, hence $\,T \ne \emptyset$.
Even more,  for any $\,p \in T\,$ it holds
$$
\# T = \# \{ p' \in T \mid p' > p\}+\# \{ p' \in T \mid p' < p\}+1,
$$
where the second member corresponds to the number of tiers above and below $\,T_p$, plus 1 for $T_p$ itself.
\end{remark}

\begin{definition} \label{def:dense rank}
Given $\,R\in \mathcal{W}(X)\,$ and $\,x_i\in X$, if $\,x_i \in T_p$, the \emph{dense rank} of $x_i$ in $R$ is defined as
\begin{equation} \label{eq:dense rank}
D_R(x_i) = \# T - \# \{ p' \in T \mid p' < p\} = \# \{ p' \in T \mid p' > p\}+1.
\end{equation}
\end{definition}

\begin{example}  \rm
Given the following weak order $\,R\in \mathcal{W}(\{x_1,\dots,x_{10}\})$
$$
\begin{array}{c}
x_3 \\[.5ex]
x_6 \;\; x_8 \\[.5ex]
x_1\;\;x_4\;\;x_7\;\;x_{10} \\[.5ex]
x_2 \;\; x_5 \;\; x_9\\
\end{array}
$$
taking into account Eqs.~(\ref{eq:Tp}), (\ref{eq:T}) and (\ref{eq:dense rank}), we have
$\,T_9 = \{ x_3\}$, $\,T_7 = \{ x_6, x_8\}$, $\,T_3 = \{ x_1, x_4, x_7, x_{10} \}$, $\,T_0 = \{ x_2, x_5, x_9 \}$,
$\,T_8=T_6=T_5=T_4=T_2=T_1=\emptyset$, $\,T = \{0,3,7,9\}\,$ (hence, $\,\#T=4$) and
\begin{eqnarray*}
&& D_R(x_3) = 4-3=1\\
&& D_R(x_6) = D_R(x_8) = 4-2=2\\
&& D_R(x_5) = D_R(x_4) = D_R(x_7) = D_R(x_{10}) = 4-1=3\\
&& D_R(x_2) = D_R(x_5) =  D_R(x_9) = 4-0=4.
\end{eqnarray*}
\end{example}

We now show that indifferent alternatives always belong to the same tier.

\begin{proposition} \label{prop:IndTier}
Given $\,R\in \mathcal{W}(X)$, for all $\,x_i , x_j \in X\,$ it holds
$$
x_i\,I\,x_j \;\Leftrightarrow\; x_i,x_j \in T_p \; \mbox{ for some }\; p\in T.
$$
\end{proposition}

\proof
$\Rightarrow$)
As $\,x_i\,I\,x_j$, if $\,x_j\,P\,x_k$, we have $\,x_i\,P\,x_k$; hence, those alternatives dominated by $\,x_j\,$ also are by $\,x_i\,$ and, consequently, $\,p_j\leqslant p_i$.
Interchanging the roles of the indifferent alternatives, from $\,x_j\,I\,x_i$, if $\,x_i\,P\,x_k$,
we obtain $\,x_j\,P\,x_k\,$ and, again in a similar way, $\,p_i\leqslant p_j$.
All in all, the alternatives dominated by $x_i$ coincide with those dominated by $x_j$.
Thus, $\,p_i=p_j\,$ and $\,x_i , x_j \in T_p$, where $\,p=p_i=p_j$.

$\Leftarrow$)
Suppose, by way of contradiction, that $\, x_i, x_j \in T_p\,$ for some $\,p\in T$, i.e., $\,p_i=p_j=p$, and not $\,x_i \,I\, x_j$.
If $\,x_i \,P\, x_j$, then $\,p_i>p_j$; and if $\,x_j \,P\, x_i$, then $\,p_j>p_i$.
\quad\qed

While tiers consider alternatives indifferent to others in an horizontal way, next we introduce vertical arrangements, from top to bottom, corresponding to different alternatives in each tier, through the notion of maximal chain, similar to that of saturated chain in the setting of partial ordered sets (see, for instance, Anderson~\cite[p. 14]{Anderson-2002}).

\begin{definition}
Given $\,R\in \mathcal{W}(X)$, a \emph{maximal $P$-chain} in $R$ is any list of alternatives $\, x_{i_1},\dots, x_{i_r} \in X \,$ such that $\, x_{i_1} \,P\, \cdots \,P\, x_{i_r}\,$ with maximum length $\,r=\# T$.
\end{definition}

\begin{remark} \label{rem:lenght-mchain} \rm
The existence of maximal $P$-chains is guaranteed being $X$ finite.
And, in fact, the maximum length ought to be $\,r=\# T$:
indeed, just selecting  one alternative in each tier $\,T_p$, with $p\in T$, we have $\,r\ge\# T$;
and such value cannot be surpassed, because if $\,r>\# T$, by the \textit{pigeonhole principle}\footnote{Also known as \textit{Dirichlet's box principle}, it states that, when $t$ objects are distributed among $\,s<t\,$ boxes (pigeonholes), at least one of them must contain at least $\,t-s\,$ objects.
In other words, there is not any bijective map of a set of cardinality $t$ into a set of cardinality $s<t$.}, at least two alternatives should lay in the same tier and, by Proposition~\ref{prop:IndTier}, they will be indifferent to each other.

Note that if $\,x_{i_l}\,$ is the $l$-th element in a maximal $P$-chain, then there should be $\,l-1\,$ tiers above, and hence  $\,D_R(x_{i_l})=(l-1)+1=l$, according to Definition~\ref{def:dense rank}.
In this way, the dense rank can also be alternatively obtained through any maximal $P$-chain as $\,D_R(x_{i_l})=l\,$ for any $\,x_{i_l}\,$ in the chain and, again by Proposition~\ref{prop:IndTier}, extending these positions to all the alternatives that are indifferent to $\,x_{i_l}$, tier by tier.
In other words, the dense rank is univocally determined by one representative element (precisely that in the maximal $P$-chain) for each equivalence class of indifferent alternatives (or tiers). 	
\end{remark}	

\section{Basic properties} \label{sect:basic properties}
We now consider some basic properties that position operators on weak orders might (or should) verify.
Note that we do not \textit{ex ante} impose compelling requirements as to assign the same positions to indifferent alternatives, etc.

\begin{definition} \label{def:position operator properties}
Let $O$ be a position operator and $\,O_R: X \longrightarrow \mathbb{R}\,$ the function that assigns a position to each alternative of $X$ in the weak order $\,R\in \mathcal{W}(X)$.
We say that the position operator $O$ satisfy the following conditions, when they are fulfilled for all $\,X \subseteq U\,$ and $\,R \in \mathcal{W}(X)$:

\begin{enumerate}
\item
\emph{Equality}: $x_i\,I\,x_j \;\Rightarrow\; O_R(x_i) = O_R(x_j)  $, for all $\,x_i,x_j\in X$.

\smallbreak
\item
\emph{Neutrality}: $O_{R^{\sigma}}(x_{\sigma(i)})=O_R(x_i)$ for every permutation $\sigma$ on $\,\{1,\dots,n\}$.

\smallbreak
\item
\emph{Sequentiality}: If $\,R\in \mathcal{L}(X)$, then $\,O_R(x_i)  = S_R(x_i)$, for every $\,x_i\in X$.

\smallbreak
\item
\emph{Truncation}: $O_{R\vert _{X\setminus T_0}}(x_i)  = O_R(x_i)$, for every $\,x_i\in X\setminus T_0$.
\end{enumerate}
\end{definition}

\begin{remark} \label{rem:truncation}  \rm
Equality entails that indifferent alternatives are indistinguishable from a positional point of view, while neutrality guarantees an equal treatment of alternatives.
Sequentiality formalizes the convention of assigning unit-equidistant positions starting by one if there are no ties.
Truncation means that the deletion of the bottom tier preserves the positions of all the remaining alternatives from the original situation.
But, by an iterative process, withdrawing the bottom tier in each step, truncation can also be equivalently formulated stating that positions of alternatives at any tier do not depend on those other alternatives staying at tiers below.
And it is important to note that this property entails that not only removing, but also adding tiers below the original situation will not change the positions of the already existing alternatives, because by successive deletions we could obtain again the situation before the expansion below.
\end{remark}

Although equality and neutrality are conditions considered from different scenarios, we now justify that the second is stronger than the first one.

\begin{proposition} \label{prop:neutrality implies equality}
If a position operator satisfies neutrality, then it also satisfies equality.
\end{proposition}

\proof
Let $O$ be a position operator satisfying neutrality and $\,R\in \mathcal{W}(X)$.
Consider $\,x_i \,I\, x_j\,$ and let $\sigma$ be  the transposition $(i,j)$.
Then, by neutrality, $\,O_R(x_i)=O_{R^{\sigma}}(x_{\sigma(i)})=O_{R^{\sigma}}(x_j)$.
Since $\,R=R^{\sigma}$ (due to the symmetry of $I$), we have $\,O_{R}(x_i)=O_{R}(x_j)$.
\quad\qed

\begin{remark} \label{rem:properties}  \rm
Standard, modified, fractional and dense ranks satisfy equality, neutrality, sequentiality and truncation.
This is straightforward for the first three properties and all these position operators due to the way they have been defined by extending the unique possibility for the linear case in different ways for tied alternatives.

Truncation fulfillment requires a more detailed explanation.
First of all, as commented in Remark~\ref{rem:truncation}, truncation implies that deletion or expansion of tiers below than that considered to assign positions to its alternatives do not affect such positions.
And this is precisely what happens with standard, modified and fractional ranks:
when assigning positions to the alternatives, only tied ones at the same tier (if any) are involved in the tie-breaking process (if needed), and not those ones below.
As for the dense rank, truncation easily follows from Definition~\ref{def:dense rank} (see second identity in Eq.~(\ref{eq:dense rank})).
\end{remark}

Consequently, as interested in characterization results, those properties in Definition~\ref{def:position operator properties} will not be selective enough to our purposes, and we have to analyze further appropriate conditions capturing the very essence of the dense rank.

\section{The characterizations} \label{sect:characterizations}
The first requirement in characterizing the dense rank will be sequentiality, which formulates a convention of equidistance between consecutive tiers, extending the case of linear orders, where tiers are singletons.
While this is not a very demanding condition, that is not the case for the other properties to appear in Theorems~\ref{th:char1} and \ref{th:char2}, duplication and UD-independency (jointly with truncation in the last case), respectively, which need more detailed explanations.

The key idea of duplication is that the process of replicating some alternatives (the pattern sources) keeps not only their original positions unchanged, but also those of the rest of non-cloned alternatives.
We have already pointed out how this condition  makes sense in some scenarios where agent's petitions can be (strategically or not) replicated.
This is the reason why this kind of ``twinning" or ``clone irrelevance" has been successfully introduced in several disciplines.
Concretely in Social Choice it appears as ``duplication" for characterizing some voting rules (see Garc\'ia-Lapresta and Mart\'inez-Panero~\cite{Lapresta-Panero-2017} and the references therein, mainly Congar and Merlin~\cite{Congar-Merlin-2012}).
The main difference is that in the voting context the entire preference relation of one agent is cloned, while in our scenario just one alternative is replicated.
Because of this analogy, we will also use the coined term ``duplication" in what follows.

A cloned alternative can be understood as an added one to those in $X$ maintaining exactly the same preference relationship than the original pattern with all the alternatives in $X$.
Hence, \textit{a fortiori}, the clone and its pattern will be indifferent to each other in the extended structure, and they are supposed to share the same position.

Similarly to what happens in the Social Choice context to deal with a variable electorate (for instance, as mentioned, when clones or new voters appear from a previous situation, as in Smith~\cite{Smith-1973}), we have also introduced the notion of position operator (Definition~\ref{def:position operator}) to tackle a similar case concerning just duplication of alternatives.

\begin{definition} \label{def:duplication}
A position operator $O$ satisfies \emph{duplication} if whenever $\,R\in \mathcal{W}(X)$, $\,R' \in \mathcal{W}(X')$,
with  $\,X' = X \cup \{ x_{n+1} \}$ such that $\,x_{n+1} \notin X$, $\,R' \vert_X = R\,$
and $\,x_{n+1} \,I'\, x_j\,$ for some $\,x_j \in X$, then $\,O_{R'} (x_i) = O_R (x_i)\,$ for every $\,x_i \in X\,$ and $\,O_{R'}(x_{n+1}) = O_{R'}(x_j)$.
\end{definition}

\begin{remark} \label{rem:indif-same} \rm
Note that $R'$ is well defined in $X'$, because from the indifference between $\,x_{n+1} \,$ and $\, x_j$, we can obtain the relationship among $\,x_{n+1} \,$ and the rest of alternatives in $X$.
\end{remark}

\begin{remark} \label{rem:dup-add-delete} \rm
Duplication means that the addition of new alternatives to an existing tier preserves the positions of all the alternatives in the original situation.
But it also entails that the deletion of alternatives in a tier does not change the positions of the remaining ones either, whenever along the withdrawal process at least one alternative (the pattern) stays in the tier.
Because, if so, by successive replication of the remaining one with the deleted alternatives, we can obtain again the situation before the deletion. Along this second stage, by duplication, positions of non-duplicated alternatives are preserved, and consequently they could not change along the previous deletion.
\end{remark}

Next we show that duplication is supported by the fact that it implies the compelling requirement of neutrality.
The outline of the proof is as follows\footnote{A similar argument appears in Congar and Merlin~\cite{Congar-Merlin-2012} within a voting framework.}:
somehow, each $\,x_{\sigma (i)}\,$ is a clone of the corresponding $\,x_i$, but we cannot directly make such association because $\,x_{\sigma (i)}\,$ already belongs to $\,X$.
This is the reason why we will introduce new alternatives and denote by $\,R^{n)}\,$ the weak order on $\,X \cup \{x_{n+1}, \dots, x_{2n} \}\,$ obtained by iterative duplication of all the alternatives in $X$, one by one, replicating each $x_j$ with $\,x_{n+j}\notin X\,$ for $\,j=1,\dots,n$.

\begin{proposition} \label{prop:dup-neu}
If a position operator satisfies duplication, then it also satisfies neutrality.	
\end{proposition}

\proof
Let $O$ be a position operator verifying duplication, $\,X=\{x_1,\dots,x_n\}$, $\,R\in \mathcal{W}(X)\,$ and a permutation $\sigma$ on $\,\{1,\dots,n\}\,$ which induces $\,R^{\sigma}\in \mathcal{W}(X)$.
We have to prove $\,O_{R^{\sigma}}(x_{\sigma (i)})=O_R(x_i)\,$ for every $\,i\in \{1,\dots,n\}$.
To this aim, duplicate every $\,x_{\sigma (i)}\,$ with $\,y_{\sigma (i)}\,$ from a set $Y$ such that $\,\#Y=\#X\,$ and $\,Y\cap X=\emptyset$, and let $\,(R^{\sigma})^{n)}$ the corresponding weak order on $\,X\cup Y$, obtained by iterated replications.
Note that, by duplication, $\,O_{(R^{\sigma})^{n)}}(x_{\sigma (i)}) = O_{(R^{\sigma})^{n)}}(y_{\sigma (i)})$.
But also each $\,y_{\sigma (i)}\,$ duplicates the original $x_i$, so that $\,O_{R^{n)}}(y_{\sigma (i)})=O_{R^{n)}}(x_i)$.
Being $\,y_{\sigma (i)}\,$ a clone of both $\,x_{\sigma(i)}\,$ and $x_i$ (\textit{caveat}:
in different extensions coinciding when restricted to $Y$), necessarily both the restrictions of $\,(R^{\sigma})^{n)}$ and $\,R^{n)}$ to $X$, which are $R^\sigma$ and $R$, respectively, must provide the same positions to the respective patterns (see Remark~\ref{rem:dup-add-delete}).
Hence, $\,O_{R^{\sigma}}(x_{\sigma (i)})=O_R(x_i)$.
\quad\qed

\begin{remark} \label{rem:same positions tier} \rm
Note that, although duplication only explicitly imposes that the pattern and its clone will have the same positions, as neutrality implies equality (Proposition~\ref{prop:neutrality implies equality}), in fact, this is also necessarily true for all the alternatives laying in the same tier.
\end{remark}

We now present the first characterization theorem.

\begin{theorem} \label{th:char1}
A position operator $O$ satisfies sequentiality and duplication if and only if for each $\,X\subseteq U\,$ finite and $\,R\in \mathcal{W}(X)$, the function $\,O_R:X \longrightarrow \mathbb{R}\,$ assigns to each $\,x_i \in X$ the dense rank of $x_i$ in $R$.
\end{theorem}

\proof
As pointed out in Remark~\ref{rem:properties}, the dense rank satisfies sequentiality due to the way it has been defined, essentially extending the linear case, where tiers are singletons and this property holds, to weak orders were tiers might have a greater cardinality.
On the other hand, after replicating an alternative from $R$ to $R'$, already existing tiers are maintained (taking into account Definition~\ref{def:dense rank} before and after the duplication process), so that the position for every alternative does not change and duplication is also satisfied.

Conversely, assume both sequentiality and duplication.
By concatenation of Propositions~\ref{prop:dup-neu} and \ref{prop:neutrality implies equality}, all the alternatives in the same tier have the same position (see Remark~\ref{rem:same positions tier}).
Let us now select just one alternative in each tier and withdraw those indifferent to them in a finite number of steps.
Then, a maximal $P$-chain will be obtained and positions will not change, as pointed out in Remark~\ref{rem:indif-same}.
In other words, if $\,C=\{ x_{i_1},\dots, x_{i_r}\}\,$ is the set of alternatives involved in a $P$-chain, then $\,R\vert_C\,$ is a linear order.
Now, by sequentiality, the positions of such alternatives ought to follow the list of the natural numbers from top to bottom.
And finally, again by successive duplication to recover the original $\,R\in \mathcal{W}(X)\,$ and taking into account the second paragraph of Remark~\ref{rem:lenght-mchain}, this exactly corresponds to the dense rank.
\quad\qed

\begin{proposition} \label{pr:independ_char1}
Sequentiality and duplication are independent.	
\end{proposition}

\proof
\begin{enumerate}
\item
Let $Q$ be the position operator that assigns to each alternative the quotient between its dense rank and the cardinality of the corresponding tier, i.e.,
\begin{equation} \label{eq:q dense rank}
Q_R(x_i) = \frac{D_R(x_i)}{\# T_p},
\end{equation}
where $\,x_i \in T_p\,$ (see Eq.~(\ref{eq:Tp})).

$Q$ verifies sequentiality, because if $\,R\in \mathcal{L}(X)$, then $\,\# T_p=1\,$ and $\,Q_R=D_R$, which satisfies this property (see Remark~\ref{rem:properties}).
However, $Q$ does not fulfil duplication: those positions of  alternatives $\,x_i \in T_p\,$ change after adding a clone, because the denominator of Eq.~(\ref{eq:q dense rank}) increases one unit.

\medbreak
\item
Let $F\,$ be the position operator (suggested by Fishburn~\cite[pp. 164-165]{Fishburn-1973}) that assigns an affine linear transformation $f$ of the dense rank:
\begin{equation} \label{eq:f dense rank}
F_R(x_i)= f\big( D_R(x_i)\big),
\end{equation}
with  $\,f(r)=ar+b\,$ and $\,a,b\geqslant 0$.

$F\,$ satisfies duplication, because the position operator that defines the dense rank does, as Theorem~\ref{th:char1} asserts.
However, $F\,$ only fulfils sequentiality whenever $f$ is the identity function in Eq.~(\ref{eq:f dense rank}), because any other affine function will expand or contract the distances among contiguous positions, or will change 1 as the starting value of the dense rank.
In particular, when $f$ is a constant function (as if all possible positions collapse in a common value), duplication trivially holds, but it is obvious that this is not true for sequentiality.
\quad\qed
\end{enumerate}

We now motivate the main property appearing in the second characterization result (jointly with truncation and sequentiality, already introduced).
This new condition entails that vertical displacements of one alternative from/to already existing tiers will not change all other alternative's positions, provided that existing tiers do not disappear and new tiers are created neither.
As happening with duplication, this requirement makes sense in some contexts.
For instance, a wage promotion (or salary degradation) should not harm (or benefit) other workers' labor conditions (this example is related to the query about the top 3 salaries in Kyte~\cite[pp. 562-568]{Kyte-2005} mentioned before).

\begin{definition} \label{def:UD-i}
A position operator $O$ satisfies \emph{UD-independency}  (UD standing for \emph{upwards or downwards}) if for all $\,R,R'\in \mathcal{W}(X)\,$ and $\,x_i,x_k,x_l\in X\,$ such that $\,x_k\,P\,x_l$,  $\,x_l\,I\,x_i$,   $\,x_k\,I'\,x_i$, $\,x_k\,P'\,x_l\,$ and $\,R' \vert_{X\setminus \{x_i\}} = R\vert_{X\setminus \{x_i\}}$,
	then $\,O_{R'}(x_j) = O_R (x_j)\,$ for every $\,x_j \in X\setminus \{x_i\}$.
\end{definition}

\begin{remark} \label{rem:UD-equivalence} \rm
UD-independency means that if an alternative jumps upwards to another existing tier (i.e., the movement does not create a new tier) while the departure tier still exists (i.e, it does not become empty), then the positions of all other alternatives remaining in the original situation are preserved.
But it also entails that the same is true for a downwards movement under the same conditions.
Because then, we can replicate the situation before by making the opposite upwards movement.
Along this second stage, by UD-independency, positions of non-moving alternatives are preserved, and consequently they could not change along the previous downwards movement.
All in all, this property can be dynamically understood as if $x_i$ were vertically moving, no matter if upwards or downwards, and note that it does not require anything to its former and current positions along the process. What we demand is that all other alternatives do not change their positions\footnote{In particular, UD-independency allows to improve an object (individual, alternative, etc.) without making any other one worse off, which is somehow related (in other context) to the notion of Pareto improvement. }.
\end{remark}

We have already pointed out that duplication and UD-independency focus on different approaches: the first one, on horizontal changes (extensions or deletions); while the second one, on vertical displacements.
Nonetheless, the following proposition justifies that duplication is a stronger condition than UD-independency.

\begin{proposition} \label{prop:dup-UP-i}
If a position operator satisfies duplication, then it also satisfies UD-independency.	
\end{proposition}

\proof
Let $O$ be a position operator, $\,R,R'\in \mathcal{W}(X)\,$ and $\,x_i,x_k,x_l\in X\,$ such that $\,x_k\,P\,x_l$,  $\,x_l\,I\,x_i$,  $\,x_k\,I'\,x_i$, $\,x_k\,P'\,x_l\,$ and $\,R' \vert_{X\setminus \{x_i\}} = R\vert_{X\setminus \{x_i\}}$.
First, withdraw $\,x_i\,$ by restricting $R$ to $\,X\setminus \{x_i\}\,$ and then extend $\,R\vert_{X\setminus \{x_i\}}\,$ by replicating $\,x_k\in X\setminus \{x_i\}\,$ with $\,x_i\,$ itself.
As the pattern $\,x_k\,$ ought to be indifferent to the clone $\,x_i\,$ in such extension of $\,R\vert_{X\setminus \{x_i\}}$, it results to be precisely $R'$.
Now, duplication implies that, along the two steps of deletion (see Remark~\ref{rem:indif-same}) and restoration (at a different tier) of $x_i$,  all the positions of alternatives other than $\,x_i\,$ are preserved, so that $\,O_{R'}(x_j) = O_R (x_j)\,$ for every $\,x_j \in X\setminus \{x_i\}$.
\quad\qed

The conditions to appear in the second characterization of the dense rank, although apparently disconnected with equality, all three together imply this last property.
It is justified in the following proposition.

\begin{proposition} \label{prop:UP-i-seq-trun-stab}
If a position operator satisfies sequentiality, truncation and UD-independency, then it also satisfies equality.	
\end{proposition}

\proof
By way of contradiction, suppose that a position operator $O$ does not satisfy equality.
Then there must exist $\,R\in \mathcal{W}(X)\,$ and two alternatives $\,x_i, x_j\in X\,$ such that $\,x_i \,I\, x_j$, $\,O_R(x_i)=r\,$ and $\,O_R (x_j)=s$, with $\,r\ne s$.
Now, take $\,x_{n+1}\notin X\,$ and extend $R$ to $R'\in \mathcal{W}(X \cup \{x_{n+1}\})\,$ by imposing $\, x_k \,P'\, x_{n+1}\,$ for all $\, x_k \in X\,$ (in other words, add a new bottom tier to those in the original situation with $\,x_{n+1}\,$ as unique alternative).
By truncation (see Remark~\ref{rem:truncation}), still $\,O_{R'}(x_j)=r\,$ and $\,O_{R'} (x_j)=s$.
Next, we select a maximal $P'$-chain $C$ containing both $\,x_i\,$ and $\,x_{n+1}$, i.e., $\, x_{i_1} \,P'\, \cdots \,P'\,x_{i} \,P'\, \cdots \,P' \,x_{n+1}$.
By applying UD-independency iteratively, we can arrange all the alternatives not appearing in $C$ as indifferent to $\,x_{n+1}\,$ (see Remark~\ref{rem:UD-equivalence}) obtaining a new weak order $\,R^*\in \mathcal{W}(X\cup\{x_{n+1}\})\,$ such that $\,R^*\vert_C = R'\vert_C$, and note that $\,O_{R^*}(x_i)=r\,$ yet.
Truncating the bottom tier of $\,R^*$ (that which gathers tied alternatives in $R'$) we obtain the linear order $\,R^*\vert_{C\setminus \{x_{n+1}\}}$, and again $\,O_{{R^*}\vert_{C\setminus \{x_{n+1}\}}}(x_i)=r$, being $\,r\in \mathbb{N}\,$ by sequentiality.

We undo all this process to recover the original $\,R\in \mathcal{W}(X)\,$ with $r$ as position of $\,x_i\,$ remaining again in each step.
And then all is repeated with $\,x_j\,$ instead of $\,x_i$,  obtaining that also necessarily  $\,s\in \mathbb{N}\,$ due to the same reasons.
Even more, as $\,x_i \,I\, x_j$, they belong to the same tier in $R$ and they share the same number $t$ of tiers above, if any (otherwise, $\,t=0$).
Hence, $\,x_i\,$ and $\,x_j$ have the same number $t$ of alternatives before in their respective maximal chains in the parallel processes.
Consequently, $\,r=t+1=s$, contrary to our assumption.
\quad\qed

We now present the second characterization theorem.

\begin{theorem} \label{th:char2}
A position operator $O$ satisfies sequentiality, truncation and UD-independency if and only if for each $\,X\subseteq U\,$ finite and $\,R\in \mathcal{W}(X)$, the function $\,O_R:X \longrightarrow \mathbb{R}\,$ assigns to each $\,x_i \in X$ the dense rank of $x_i$ in $R$.
\end{theorem}

\proof
As sequentiality is a common condition in both characterization results, its fulfilment by the dense rank was already justified in the proof of Theorem~\ref{th:char1}.
The dense rank also satisfies truncation, as pointed out in Remark~\ref{rem:properties}.
And, in order to complete necessity, notice that along the process followed in Definition~\ref{def:UD-i}, the moving alternative $\,x_i\,$ is displaced from an existing tier to another existing one, so that according to Definition~\ref{def:UD-i}, those alternatives in $X$ other than $\,x_k\,$ will not change their dense rank positions, which is precisely what UD-independency requires.

For sufficiency, suppose that a position operator $O$ satisfies simultaneously the three above mentioned properties, and let us apply them to $\,R\in \mathcal{W}(X)\,$ for showing that $\,O_R=D_R$.
First, select a maximal $P$-chain $C$ in $\,R\,$ given by $\, x_{i_1} \,P\, \cdots \,P \,x_{i_l}\,$ and  take $\,x_{n+1}\notin X\,$  extending $R$ to $R'\in \mathcal{W}(X \cup \{x_{n+1}\})\,$ by imposing $\, x_{i_l} \,P'\, x_{n+1}$, so that $\,C\cup \{x_{n+1}\}\,$ will also be a maximal $P'$-chain and, by truncation, $\,O_{R'}(x_{i_m})=O_R(x_{i_m})\,$ for each $\,x_{i_m}\in C$.

Now, the following part of the current proof is similar to that of Proposition~\ref{prop:UP-i-seq-trun-stab}, playing also here this last new tier a pivotal role, as follows.
By applying UD-independency iteratively, we can arrange all the alternatives in $\,X\setminus C\,$ as indifferent to $\,x_{n+1}\,$ in a new weak order $\,R^*\in \mathcal{W}(X\cup\{x_{n+1}\})\,$ which coincides with $R'$ (and hence with $R$) over $C$.
Thus, $\,O_{R^*}(x_{i_m})=O_{R'}(x_{i_m})=O_R(x_{i_m})\,$ for each $\,x_{i_m}\in C$.
Truncating the bottom tier of $\,R^*\,$ (that which gathers tied alternatives in $R$), we obtain the linear order $\,{R^*}\vert_C\,$ and again $\,O_{{R^*}\vert_C}(x_{i_m})=O_{R^*}(x_{i_m})=O_{R'}(x_{i_m})=O_R(x_{i_m})\,$ for each $\,x_{i_m}\in C$.
Now, by sequentiality in $\,{R^*}\vert_C$, we have  $\,O_{{R^*}\vert_C}(x_{i_m})=O_R(x_{i_m})=m$, i.e., the alternatives $\, x_{i_1},  \dots ,\, \,x_{i_l} \in C\,$ have original positions $\,1,\dots,l$, respectively.
Finally, undoing the previous process for recovering $R$, each element in $C$ will determine the positions of all the alternatives in its tier, by Proposition~\ref{prop:UP-i-seq-trun-stab}.
Hence, we obtain that $\,O_R=D_R\,$ (see the second paragraph of Remark~\ref{rem:lenght-mchain}), in a similar way to what happened at the end of the proof of Theorem~\ref{th:char1}.
\quad\qed

Once this second characterization has been achieved, we now justify the independence of the three conditions appearing in Theorem~\ref{th:char2}.

\begin{proposition} \label{pr:independ_char2}
Sequentiality, truncation and UD-independency are independent.
\end{proposition}

\proof
\begin{enumerate}
\item
The position operator defined by $Q_R\,$ in Eq.~(\ref{eq:q dense rank}) satisfies sequentiality and truncation, but not UD-independency.

Sequentiality of $Q_R\,$ has been already proven in Proposition \ref{pr:independ_char1}.
$Q_R\,$ also fulfils truncation, because the numerator $\,D_R(x_i)\,$ does (by Remark~\ref{rem:properties}), and the denominator is not affected by a deleted tier below.
However, the position operator defined by $Q_R$ does not satisfy UD-independency, because now, along the $\,x_i\,$ moving process in Definition~\ref{def:UD-i}, the cardinality of the departure and arrival tiers change, and hence also do those positions of all the alternatives in both tiers.

\medbreak
\item
The position operator defined as
$$
O_R(x_i)= \left\{ \begin{array}{ll}
D_R(x_i)+n, & \mbox{ if } \, R \in \big(\mathcal{W}(X) \setminus \mathcal{L}(X)\big),\\[.5ex]
D_R(x_i), & \mbox{ if } \, R \in \mathcal{L} (X)
\end{array}\right.
$$
satisfies sequentiality and UD-independency but not truncation.

By definition, if $\,R \in \mathcal{L} (X)$, then $O_R=D_R$, which satisfies sequentiality (see Remark~\ref{rem:properties}).
As UD-independency is not intended for linear orders (because Definition~\ref{def:UD-i} requires indifference between different alternatives), we use $\,O_R(x_i)=D_R(x_i)+n\,$ and this property holds, inherited from that of $\,D_R\,$ and taking into account that $n$ remains constant along the vertical moving process.
However, a weak order can become linear by truncation, and in such case positions of the remaining alternatives will change.

\medbreak
\item
The position operator defined as $\,O_R(x_i)=i\,$ satisfies truncation and UD-independency but not sequentiality.

In fact, this position operator does not depend on the relationship among alternatives, but just on their list order.
Hence, truncation is satisfied because, after deleting some bottom alternatives, those remaining will keep their list numbers.
The same holds for UD-independency along the moving process.
However, sequentiality does not hold because the list order might not coincide with the sequential positions given by the linear order.
\quad\qed
\end{enumerate}

\begin{remark} \rm
Next, discarding in our analysis the common property considered in both characterizations (sequentiality), we will justify that duplication (appearing in Theorem~\ref{th:char1}) is not decomposable in  UD-independency jointly with truncation (appearing in Theorem~\ref{th:char2}).
Otherwise, both given axiomatizations  would be essentially the same; but this is not the case, and Theorem~\ref{th:char2} is neither equivalent nor a consequence of Theorem~\ref{th:char1}.
To show this, we will demonstrate that although duplication implies UD-independency (Proposition~\ref{prop:dup-UP-i}), it does not entail truncation.
Consider, for instance, the position operator defined as
$$
O_R(x_i)= \frac{D_R(x_i)}{\# T}.
$$

This position operator satisfies duplication, because it is a particular case of that already defined in Eq.~(\ref{eq:f dense rank}) verifying this property, by simply taking $f$ a contraction (dividing by $\,\# T$).
But this denominator is precisely which prevents $\,O_R\,$ from truncation fulfilment, because after deleting the last tier, $\,\# T\,$ changes (it is reduced one unit), and hence $\,O_R(x_i)\,$ also does.

Additionally, the relationships of both duplication and the combination of UD-independency plus truncation with neutrality (interesting by themselves) sheds some more light on the above comments on the achieved characterizations, and confirm that they are essentially different.
Indeed, while duplication implies neutrality (Proposition~\ref{prop:dup-neu}), we will show that UD-independency jointly with truncation do not.
Consider again the position operator given by $\,O_R(x_i)=i\,$, which satisfies UD-independency and truncation, as justified before.
However, it is easy to check that it does not fulfil equality (indifferent alternatives have distinct list numbers) and, by Proposition~\ref{prop:neutrality implies equality}, it does not satisfy neutrality either.
\end{remark}

\begin{remark} \label{rem:implications}  \rm
According to Propositions~\ref{prop:dup-neu} and \ref{prop:neutrality implies equality}, duplication implies neutrality, and neutrality implies equality, respectively.
Hence, due to Theorems~\ref{th:char1} and \ref{th:char2}, the position operator that defines the dense rank satisfies all the properties appearing in the paper: equality, neutrality, sequentiality, truncation, duplication and UD-independency.
Note that the first four mentioned conditions are also satisfied by standard, modified and fractional ranks, while duplication and UD-independency are specific of the dense rank.
\end{remark}

\begin{center}
\begin{figure}[h]
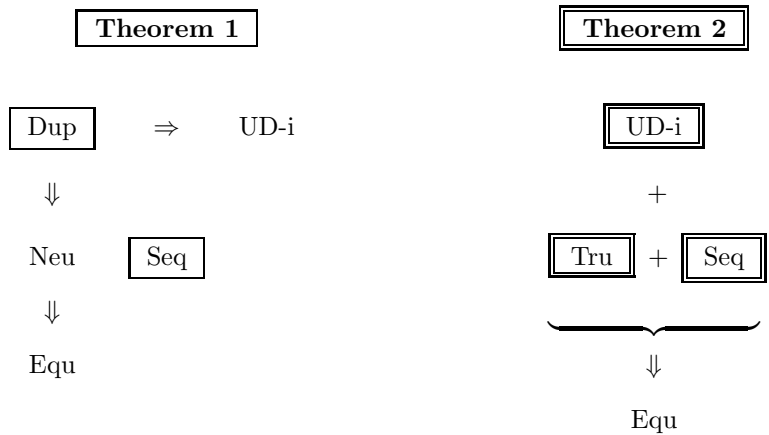

\centering
$$
\begin{array}{cccc}
& \fbox{\textbf{ Theorem \ref{th:char1}} } & & \qquad\qquad\qquad\qquad \doublebox{\textbf{ Theorem \ref{th:char2}} } \\[5ex]
\fbox{ Dup } \hspace{-6mm} & \Rightarrow & \hspace{-6mm} \mbox{UD-i } & \qquad\qquad\qquad\qquad \doublebox{ UD-i }\\[3ex]
 \Downarrow \hspace{-6mm} & & \hspace{-6mm} & \qquad\qquad\qquad\qquad\;  +\\[2ex]
 \mbox{Neu}  \hspace{-6mm} & \fbox{ Seq } & \hspace{-6mm} & \qquad\qquad\qquad\qquad\; \doublebox{ Tru } \;+\; \doublebox{ Seq }\\[2ex]
 \Downarrow \hspace{-6mm} & & \hspace{-6mm} & \qquad\qquad\qquad\qquad \underbrace{\hskip 8em}\\[2ex]
 \mbox{Equ} \hspace{-6mm} & & \hspace{-6mm} & \qquad\qquad\qquad\qquad \Downarrow\\[2ex]
 \hspace{-6mm} & & \hspace{-6mm} & \qquad\qquad\qquad\qquad \mbox{Equ} \\
\end{array}
$$
\caption{\label{fig:implications}Relationships among conditions.}
\end{figure}
 \end{center}

In a synoptic way, Fig.~\ref{fig:implications} shows some relationships among the 6 conditions used in the paper (in boxes, those characterizing the dense rank in Theorems~\ref{th:char1} and \ref{th:char2}).

\section{Concluding remarks} \label{sect:concluding remarks}
Along the paper we have focused on the dense rank, but we have also dealt with the standard, modified and fractional ranks.
Concerning the duplication property, when the dense rank is used, if a clone is introduced, then the positions of all the already existing alternatives remain.
However, it is straightforward to see that, with the standard rank, only the alternatives in the same tier of the clone and those above do not change their positions, while under both modified and fractional ranks just alternatives in tiers above the clone do not change their positions.

Regarding UD-independency, it is easy to check that, when making vertical displacements of alternatives from $\,k^{th}\,$ to $\,l^{th}\,$ tiers or vice-versa, with $\,k<l$, all the alternatives staying in both tiers or between them (if any) change their positions with  fractional rank;
when using the standard rank, there are variations of the positions of those alternatives staying in, or between (if any) $\,(k+1)^{th}\,$ and $\,l^{th}\,$ tiers;
and with the modified rank, the same happens for alternatives staying in, or between (if any) $\,k^{th}\,$ and $\,(l-1)^{th}\,$ tiers.
However, with the dense rank all positions remain unchanged except those of displaced alternatives.

Other differences can be observed between the dense rank and the standard, modified and fractional ranks.
From the mere knowledge of the reached positions by a list of alternatives (but not their multiplicities), we can induce to some extent their preference/indifference arrangement and how many of them are involved in, but not which of them.
More precisely, we can obtain the arrangement of the weak order with standard, modified and fractional ranks.
However, if the dense rank is used, we can not determine the structure of the weak order, but just ensure that the cardinality of the set of alternatives will be equal than or greater to the number of tiers, and no more.
This fact emphasizes again that the dense rank essentially differs from the standard, modified and fractional ranks (duplication and UD-independency are behind this behavior).

Note that we have achieved the characterizations appearing in the paper without imposing the compelling requirement of \emph{monotonicity}:
$\,x_i\,R\,x_j \;\Leftrightarrow\; O_R(x_i) \leqslant O_R(x_j)$, a stronger condition than equality.
Of course, other axiomatizations of the dense rank would be possible by incorporating this property.

The dense rank makes sense in several contexts (human resources, management, logistics, etc.) and can be used to prevent strategies (with false-name-proof mechanisms).
Thus, we advocate for it as an appropriate position operator in some scenarios, although with a different approach than those of the standard, modified and fractional ranks.
Characterizations of these last three in this positional setting and a comprehensive framework enclosing all of them are still to be provided.

\section*{Acknowledgments}
The authors are grateful to Marina N\'u\~{n}ez and the participants to the \emph{13th Conference on Economic Design} (Girona, June 2023) for their comments and suggestions.
The financial support of the Spanish \emph{Ministerio de Ciencia e Innovaci\'on} (project PID2021-122506NB-I00) is acknowledged.

\end{document}